\documentclass[prd,aps,preprintnumbers,nofootinbib]{revtex4-1}

\newcommand{\beq}{\begin{equation}}
\newcommand{\eeq}{\end{equation}}
\newcommand{\be}{\begin{eqnarray}}
\newcommand{\ee}{\end{eqnarray}}

\usepackage{amsmath}
\usepackage{amsfonts}
\usepackage{amssymb}
\usepackage[margin=1in]{geometry}
\usepackage[multiple]{footmisc}
\usepackage{hyperref}
\usepackage{graphicx}
\usepackage{mathtools}
\usepackage{physics}
\usepackage{slashed}
\usepackage{xcolor}

\begin{document}

\preprint{APCTP Pre2024-015}

\title{\boldmath
Charged Higgs Boson Phenomenology
\\
in the Dark $Z$ mediated Fermionic Dark Matter Model
}

\author{ Kyu Jung Bae }
\email{kyujung.bae@knu.ac.kr}

\affiliation{
Department of Physics, Kyungpook National University,
Daegu 41566, Korea}

\author{ Jinn-Ouk Gong }
\email{jgong@ewha.ac.kr}

\affiliation{
Department of Science Education, Ewha Womans University,
Seoul 03760, Korea
\\
Asia Pacific Center for Theoretical Physics,
Pohang 37673, Korea}

\author{ Dong-Won Jung }
\email{dongwon.jung@ibs.re.kr}

\affiliation{
Cosmology, Gravity and Astroparticle Physics Group,
Center for Theoretical Physics of the Universe,\\
Institute for Basic Science,
Daejeon, 34126, Korea
}

\author{ Kang Young Lee }
\email{kylee.phys@gnu.ac.kr; Corresponding Author}

\author{ Chaehyun Yu }
\email{chyu@kias.re.kr}

\affiliation{
Department of Physics Education \& RINS,
Gyeongsang National University, Jinju 52828, Korea
}

\author{ Chan Beom Park }
\email{cbpark@jnu.ac.kr; Corresponding Author }

\affiliation{
Department of Physics and IUEP, Chonnam National University,
Gwangju 61186, Korea
}

\date{\today}

\begin{abstract}

We present the phenomenology of the charged Higgs boson $H^\pm$
appearing in a fermionic dark matter model
mediated by an additional scalar doublet.
In order to couple the dark matter fermion to the scalar doublet,
we introduce a U(1)$_X$ gauge symmetry,
which is spontaneously broken at electroweak symmetry breaking,
resulting in a massive $Z'$ gauge boson.
Since $Z'$ is generically light,
the model is subject to strong constraints 
from electroweak precision observables.
As a result, the charged Higgs boson mass allowed
by current experimental bounds
is typically light in this model,
110 GeV $<m_{H^\pm}<$ 170 GeV.
Such a light charged Higgs boson will be produced
mainly through top-quark decays at the LHC.
Additionally, depending on the mass of 
the additional neutral Higgs boson $h$ and the dark gauge boson $Z'$,
the direct production channels
$pp \to H^\pm Z'$ and $pp \to H^\pm h$ can become sizable.
We investigate the corresponding signal processes at the LHC
to assess the discovery potential for $H^\pm$.
Current ATLAS and CMS searches for light charged Higgs bosons already impose
further constraints on the model. We also discuss the implications of dark
matter in relation to the charged Higgs boson phenomenology.

\end{abstract}

% \pacs{ }

\maketitle

\section{Introduction}

Many theoretical models of physics beyond the Standard Model (SM)
involve extensions of the Higgs sector.
In particular, the charged Higgs boson $H^\pm$ is of great interest, as it arises naturally in many new physics scenarios but is absent in the SM.
Thus, the discovery of a charged Higgs boson would provide indisputable evidence for new physics beyond the SM.
Experimental searches for charged Higgs bosons have been extensively performed
at the CERN Large Hadron Collider (LHC) through various search
channels~\cite{ATLAS:2014otc,CMS:2015lsf,ATLAS:2018gfm,CMS:2019bfg,
CMS:2015yvc,CMS:2020osd,ATLAS:2013uxj,ATLAS:2024oqu,ATLAS:2024hya,
CMS:2018dzl,CMS:2019idx,
CMS:2022jqc,ATLAS:2015edr,CMS:2017fgp,
CMS:2021wlt,CMS:2020imj,CMS:2019rlz,ATLAS:2021upq,
ATLAS:2018ntn,ATLAS:2023bzb}.
In two-Higgs-doublet models (2HDMs) and their extensions,
the dominant decay channels of the charged Higgs boson are typically
$H^\pm \to tb$ and $H^\pm \to cs$, depending on its mass,
while $H^\pm \to \tau \nu$ can also be significant.
Since no experimental evidence for the charged Higgs boson
has been observed so far,
it remains important to explore possible collider signatures
and the related phenomenology of the charged Higgs boson
in the ongoing search for new physics beyond the SM.

In this paper,
we investigate the phenomenology and discovery potential
of the charged Higgs boson at the LHC
in a hidden sector model,
where an additional Higgs doublet serves as a messenger field
between the hidden and SM sectors\cite{Jung:2020ukk,Jung:2021bjt,Jung:2023ckm,Jung:2023doz}.
In models featuring singlet fermionic dark matter (DM), 
the hidden sector is often connected to SM particles 
by a scalar singlet
\cite{Kim:2006af,Kim:2008pp,Kim:2018ecv}. 
However, in this model, the messenger field is an SU(2) scalar doublet, 
which requires a new interaction 
for the dark matter fermion to couple with the scalar doublet.
As a minimal setup, the hidden sector is assumed to be QED-like, containing a
hidden Dirac fermion as a dark matter (DM) candidate charged under a new
U(1)$_X$ gauge symmetry.
The messenger Higgs doublet field carries the U(1)$_X$ charge
and couples to the hidden sector,
while it is forbidden from coupling to the SM fermions.
From the viewpoint of the visible sector,
the model contains two Higgs doublets,
but only one of them couples to the SM fermions,
yielding a flavor structure identical to that of a type-I 2HDM.~\cite{Ko:2012hd}
The U(1)$_X$ gauge symmetry is broken
by the vacuum expectation value (VEV) of the messenger field
at electroweak symmetry breaking (EWSB),
giving mass to an additional neutral gauge boson, $Z'$.
The $Z'$ boson mixes with the SM $Z$ boson,
and its couplings to SM fermions resemble those of the SM $Z$,
but are suppressed by the $Z$--$Z'$ mixing angle.
Following Ref.~\cite{Davoudiasl:2012ag}, we refer to this particle as the ``dark $Z$ boson.''
Consequently, the DM fermion interacts with the SM sector
via the dark $Z$ mediator,
and the model effectively becomes a hybrid
of a type-I 2HDM and a dark-$Z$ portal connecting DM to the SM.
The phenomenology of such massive dark gauge bosons
has been widely explored in the literature
as a well-motivated extension of the
SM~\cite{Jung:2020ukk,Jung:2021bjt,Jung:2023ckm,Jung:2023doz,Davoudiasl:2012ag,Davoudiasl:2012qa,darkZ2}.

After EWSB, the model contains
a charged Higgs boson pair $H^\pm$ and a neutral Higgs boson $h$
as new components of the Higgs sector.
The CP-odd neutral Higgs boson $A$ is absent
because the U(1)$_X$ symmetry is broken,
and its degree of freedom is absorbed
into the longitudinal mode of $Z'$.
Experimental constraints on this model
from electroweak precision data and Higgs-sector observables
have been studied in
Refs.~\cite{Jung:2020ukk,Jung:2023ckm,Jung:2021bjt,Jung:2023doz},
where it was found that both the $Z'$ and $H^\pm$ are typically light.
The scenario in which the charged Higgs boson can remain light
when $\tan\beta$ is sufficiently large
is one of the characteristic features of a type-I 2HDM~\cite{Mondal:2023wib,Cheung:2022ndq}.

In this paper, we update the analysis
using the latest version of the public codes
{\tt HiggsBounds}~\cite{Bechtle:2020pkv}
and {\tt HiggsSignals}~\cite{Bechtle:2020uwn},
which are encoded in {\tt HiggsTools}~\cite{Bahl:2022igd},
to determine the allowed parameter space
and explore the LHC phenomenology of $H^{\pm}$ accordingly.
We find that the window of $m_{H^\pm}$
lies between 110 GeV and 170 GeV. In this model,
such a light $H^\pm$ is produced predominantly
in top-quark decays.
Moreover,
its bosonic decay modes involving a $W^\pm$ boson
play a more important role than the conventional fermionic channels
$H^\pm \to cs$ and $H^\pm \to \tau\nu$.
The relevant bosonic modes are
$H^\pm \to W^\pm h$ and $H^\pm \to W^\pm Z$.
Additionally, direct production of $H^\pm$ in
association with $Z'$ or $h$ can be sizable.
It is therefore essential to develop strategies
to identify the light $Z'$ and $h$ at the LHC.

The DM phenomenology strongly depends
on the U(1)$_X$ charge and the mass of the DM fermion.
While the $Z'$ mass and its couplings to the SM fermions
are tightly constrained by electroweak precision measurements,
the DM-$Z'$ coupling still offers some flexibility
to adjust the DM properties relevant for cosmology and astrophysics.
Once the U(1)$_X$ charge of the DM fermion is fixed
to reproduce the correct relic abundance via freeze-out,
a large portion of the parameter space becomes disfavored
by direct detection experiments.
We find that only a very limited parameter region remains
consistent with current DM constraints.

The outline of this paper is as follows. 
In Sec.~\ref{sec:model}, we briefly review the model. 
Section~\ref{sec:constraint} presents the allowed parameter sets
based on electroweak and Higgs-sector constraints. 
In Sec.~\ref{sec:lhc}, we discuss the decays of the charged Higgs boson, 
its production mechanisms at the LHC, 
and the corresponding discovery potential. 
The DM phenomenology, in relation to the charged Higgs mass 
and mixing parameters, is described in Sec.~\ref{sec:dm}. 
Finally, our conclusions are summarized in
Sec.~\ref{sec:conc}.

\section{\boldmath Dark $Z$ mediated Fermionic Dark Matter Model}
\label{sec:model}

We consider a hidden sector that includes a SM gauge singlet
Dirac fermion $\psi_X$, charged under a new $\mathrm{U}(1)_X$ gauge symmetry.
The DM candidate $\psi_X$ is a SM singlet
fermion with gauge charges assigned as $(1, \, 1, \, 0, \, X)$ 
under the gauge group
$\mathrm{SU}(3)_c \times$ $\mathrm{SU}(2)_L \times$ $\mathrm{U}(1)_Y \times$
$\mathrm{U}(1)_X$. The SM fields are neutral under $\mathrm{U}(1)_X$ 
and do not couple directly to $\psi_X$. 
In this work, we neglect the kinetic mixing between
the hidden $\mathrm{U}(1)_X$ and the SM $\mathrm{U}(1)_Y$. 
One can find the effects of the kinetic mixing on the phenomenology 
of the similar model in Refs. \cite{Davoudiasl:2012ag,Barik:2025wnh}. 

An additional $\mathrm{SU}(2)$ scalar doublet $H_1$ is introduced as a mediator
field between the hidden sector and the SM sector. The $\mathrm{U}(1)_X$ gauge
coupling $g_X$ is a free parameter of the model, and we set the
$\mathrm{U}(1)_X$ charge of $H_1$ to be $1/2$ for convenience. 
The gauge charges of $H_1$ and the SM-like Higgs doublet $H_2$ 
are given by $(1, \, 2, \, 1/2, \, 1/2)$ and $(1, \, 2, \, 1/2, \, 0)$, 
respectively. 
Due to its $\mathrm{U}(1)_X$ charge, 
$H_1$ does not couple to the SM fermions, and
only $H_2$ couples to the SM fermions via the SM Yukawa interactions. 
Thus, the Higgs field content corresponds to that of the type-I 2HDM.

The Lagrangian for the Higgs sector is given by
\be
{\cal L}_H = (D^\mu H_1)^\dagger D_\mu H_1
	   + (D^\mu H_2)^\dagger D_\mu H_2 - V(H_1, H_2)
	   + {\cal L}_{\rm Y}(H_2),
\ee
where the covariant derivative $D^\mu$ is modified to be
$D^\mu = \partial^\mu + i g W^{\mu a} T^a
                 + i g' B^\mu Y + i g_X A_X^\mu X$
by the hidden U(1)$_X$ charge operator, $X$,
and ${\cal L}_{\rm Y}$ the Yukawa couplings.
The Higgs potential reads
\be
V(H_1,H_2) &=& \mu_1^2 H_1^\dagger H_1 + \mu_2^2 H_2^\dagger H_2
              + \lambda_1 (H_1^\dagger H_1)^2
	      + \lambda_2 (H_2^\dagger H_2)^2
\nonumber \\
	   && + \lambda_3 (H_1^\dagger H_1)(H_2^\dagger H_2)
              + \lambda_4 (H_1^\dagger H_2)(H_2^\dagger H_1).
\ee
We note that the soft $Z_2$ symmetry breaking terms $H_1^\dagger H_2$
and $(H_1^\dagger H_2)^2$
are forbidden by the U(1)$_X$ gauge symmetry.

The two Higgs doublets develop the VEVs, 
$\langle H_i \rangle = (0,v_i/\sqrt{2})^T$ with $i=1$, $2$,
which break SU(2)$_L \times$ U(1)$_Y \times$ U(1)$_X$
down to U(1)$_{\rm EM}$.
We define $\tan \beta \equiv v_2/v_1$.
After EWSB,
the neutral gauge fields $(A_X^\mu, W_3^\mu, B^\mu)$
acquire masses to be respectively
the physical gauge bosons, viz. massless photon, the ordinary $Z$
and an extra $Z$ ($Z'$) bosons such that
\be
	A_X &=& c_X Z' + s_X Z,
\nonumber \\
	W_3 &=& -s_X c_W Z' + c_X c_W Z + s_W A ,
\nonumber \\
	B &=& s_X s_W Z' - c_X s_W Z + c_W A
\ee
in terms of the weak mixing angle,
$s_W \equiv \sin \theta_W$,
and the $Z$--$Z'$ mixing angle, $s_X \equiv \sin \theta_X$,
which is defined by
\beq
\tan 2 \theta_X
                = \frac{-2 g_X g' s_W \cos^2 \beta }
		       {{g'}^2 - g_X^2 s_W^2 \cos^2 \beta}.
\eeq
Since the U(1)$_X$ symmetry is broken only by the VEV of the $H_1$, 
the mass of the $Z'$ boson is intrinsically tied to the electroweak scale. 
Without a new scale in the model,
it is natural that the new physics effect is small, 
$v_1^2 \ll v_2^2 \sim v^2$ and $\theta_X \ll 1$,
from the electroweak precision variables. 
Consequently it is favored that 
the $Z'$ mass is lighter than the $Z$ boson mass.
The neutral current (NC) interactions of $Z$ and $Z'$ bosons
are given by
\be
{\cal L}_{\rm NC} =
	 \left( c_X {Z}^\mu + s_X {Z'}^\mu \right)
	 \left( g_V \bar{f} \gamma_\mu f
	  + g_A \bar{f} \gamma_\mu \gamma^5 f \right),
\ee
where $g_V$ and $g_A$ are the SM vector and axial-vector couplings
of the $Z$ boson to the fermions.
Note that the couplings of $Z$ and $Z'$ to fermions are structurally identical,
except for the overall suppression factor $s_X$ 
associated with the $Z'$ contribution.

The physical states of the neutral and charged Higgs bosons are defined
in terms of mixing angles.
The neutral Higgs mixing angle $\alpha$ is a free parameter of the model,
while the charged Higgs mixing angle is $-\beta$.
The charged states are given by
\be
        \left( \begin{array}{c}
        H_1^\pm  \\[1pt]
        H_2^\pm  \\[1pt] \end{array}
        \right)
      = \left( \begin{array}{c}
        G^\pm \cos \beta - H^\pm \sin \beta  \\[1pt]
        G^\pm \sin \beta + H^\pm \cos \beta  \\[1pt] \end{array}
        \right),
\ee
where
the massless modes $G^\pm$ correspond to
the longitudinal components of the $W^\pm$ bosons.

\section{ Constraints }
\label{sec:constraint}

Generically, the dark $Z$ boson is lighter than 
the ordinary $Z$ boson in this model, 
which imposes strong constraints on electroweak observables. 
Since $m_Z$ is modified while $m_W$ remains unchanged, 
their tree-level relation is modified
to
\be
m_Z^2 &=& \frac{m_W^2}{c_W^2 c_X^2} - m_{Z'}^2 \frac{s_X^2}{c_X^2},
\ee
leading to a shift in the $\rho$ parameter and 
stringent constraints on the model. 
Note that there are also new scalar contributions to $\Delta\rho$ in this
setup, which we consider at one-loop level~\cite{Jung:2023ckm}.

The most stringent constraint on the extra NC interactions
arises from the precise measurement
of the atomic parity violation.
The weak charge receives a leading-order deviation in $s_X$ due to dark $Z$ exchange, given by
$Q_W = Q_W^{\rm SM} ( 1+s_X^2 m_Z^2/m_{Z'}^2)$,
which leads to the bound
\beq
\frac{m_Z^2}{m_{Z'}^2} s_X^2 \le 0.006
\eeq
at 90\% confidence level (CL), based on the current experimental
value for the cesium atom,
$ Q_W^{\rm exp} = -73.16\pm0.35$
\cite{Porsev:2010de}.
We use the SM prediction $ Q_W^{\rm SM} =  -73.16\pm0.05$
\cite{Marciano:1982mm,Marciano:1990dp}.

\begin{table}[b]
\begin{tabular}{cccccccccc}
\hline
\hline
	&&$m_{Z'}$ &&&&&  9.3 GeV $-$ 11.3 GeV&&
\\
	&&$\sin \theta_X$ &&&&&  $-0.007$ -- 0&&
\\
\hline
	&&$m_{H^\pm}$ &&&&& 110 GeV $-$ 170 GeV&&
\\
	&&$\tan \beta$ &&&&& $>$ 2.1 &&
\\
\hline
	&&$m_{h}$ &&&&&  10 GeV -- 310 GeV&&
\\
	&&$\sin \alpha$ &&&&& $-0.17$ -- 0.56 &&
\\
\hline
\hline
\end{tabular}
\caption{
Allowed parameter regions of $m_{Z'}$, $\sin\theta_X$,
$m_{H^\pm}$, $\tan \beta$, $m_h$, and $\sin \alpha$
from the electroweak and Higgs phenomenology constraints.
}
\label{tab1}
\end{table}

\begin{figure}[t]
%%\centering
\includegraphics[width=\textwidth]{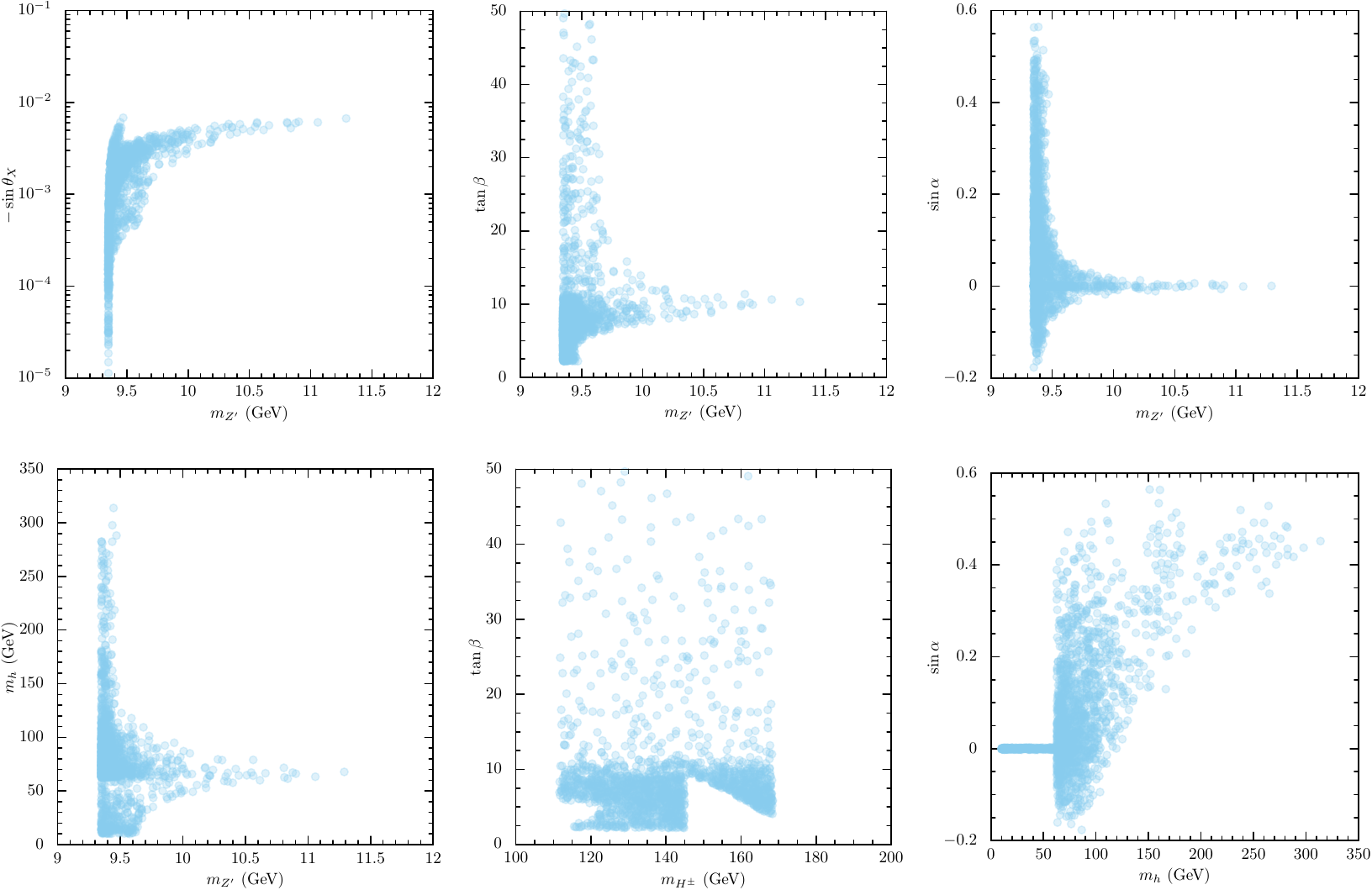}
\caption{
Parameter regions allowed by the electroweak constraints,
{\tt HiggsBounds}, and {\tt HiggsSignals}.
Shown are the viable ranges in the planes of
$(m_{Z'}, \sin \theta_X)$,
$(m_{Z'}, \tan \beta)$,
$(m_{Z'}, \sin \alpha)$,
$(m_{Z'}, m_h)$,
$(m_{H^\pm}, \tan \beta)$.
$(m_h, \sin \alpha)$.
\label{fig:allowed}
}
\end{figure}

Before applying experimental constraints to the Higgs sector, we first consider
several theoretical requirements on the Higgs potential. We demand the
perturbativity of the quartic couplings, requiring $|\lambda_i| < 4\pi$.
Additionally, the potential must be bounded from below, which imposes the
following conditions:
\be
\lambda_1 > 0,\quad
\lambda_2 > 0,\quad
\lambda_3 > -2\sqrt{\lambda_1 \lambda_2},\quad
\lambda_3 + \lambda_4 > -2\sqrt{\lambda_1 \lambda_2}.
\ee
We also require perturbative unitarity of $WW$ scattering, including
contributions from neutral scalar interactions at tree
level~\cite{Lee:1977eg,Muhlleitner:2016mzt}. The detailed conditions are
provided in Ref.~\cite{Jung:2023ckm}.

By diagonalizing the mass matrices, we obtain two CP-even neutral Higgs bosons,
$h$ and $H$, and a pair of charged Higgs bosons $H^\pm$ as physical states. We
fix the mass of one neutral Higgs boson to match the observed SM Higgs boson
mass, $m_H = 125~\mathrm{GeV}$~\cite{PDG}, identifying $H$ as the SM-like Higgs.
The mass of the other neutral Higgs boson $h$ remains a free parameter.
The masses of the charged states are given by $0$ and $M_\pm^2 = -\lambda_4 v^2
/ 2$, corresponding to the Goldstone mode and the physical charged Higgs boson,
respectively. The mixing angle for the neutral Higgs bosons is denoted by
$\alpha$, while the mixing angle for the charged Higgs bosons is $-\beta$.

The newly introduced model parameters include the $\mathrm{U}(1)_X$ gauge
coupling $g_X$ and additional Higgs sector parameters: $\mu_2^2$, $\lambda_2$,
$\lambda_3$, and $\lambda_4$, which appear in the visible sector of the model.
We reparametrize these into the following five independent parameters:
\begin{equation}
(m_{Z'},\, m_h,\, \sin\alpha,\, m_{H^\pm},\, \tan\beta),
\label{eq:para}
\end{equation}
although $g_X$ and the $Z$--$Z'$ mixing angle $\theta_X$ will still be
explicitly used in later expressions for clarity. The new gauge coupling is
assumed to satisfy the perturbativity condition,
${g_X^2}/ {(4\pi)} \leq 1$.

The constraints on our model from Higgs boson phenomenology are evaluated using
{\tt HiggsBounds}~\cite{Bechtle:2020pkv} and 
{\tt HiggsSignals}~\cite{Bechtle:2020uwn}, 
which are implemented in {\tt HiggsTools}~\cite{Bahl:2022igd}.
{\tt HiggsBounds} provides 95\% exclusion limits for additional scalar
production at collider experiments such as LEP, Tevatron, and the LHC.
{\tt HiggsSignals} computes the $\chi^2$ value of the model with respect to
SM-like Higgs boson observables at the LHC. We impose the condition
$\Delta\chi^2 = \chi^2 - \chi_{\rm min}^2 \leq 12.592$ 
corresponding to a 95\% CL.

When the dark $Z$ boson and
the additional neutral Higgs boson $h$ are light,
the $Z$ boson can decay into $Z' h$.
Since the total decay width of the $Z$ boson is
precisely measured by the LEP and SLC~\cite{ALEPH:2005ab}
$\Gamma_Z = 2.4955 \pm 0.0023$~GeV,
the contribution from the decay $Z \to Z' h$ to $\Gamma_Z$ provides a strong
constraint on the model parameters.

The allowed values of the model parameters given in Eq.~(\ref{eq:para}) are
presented in Fig.~\ref{fig:allowed} and summarized in Table~\ref{tab1}, 
subject to all electroweak and Higgs sector constraints. 
We find that there exists a very narrow window 
for the dark $Z$ boson mass: $9.3~\mathrm{GeV} < m_{Z'} < 11.3~\mathrm{GeV}$. 
This highly restrictive parameter space leads to definite
predictions for signal processes involving the dark $Z$ boson, 
which can be tested unambiguously in the near future~\cite{Jung:2023doz}.
We note that the parameter space shown in Fig.~\ref{fig:allowed}
involving such a light $Z'$ has been severely tested against LEP searches 
for charged and neutral Higgs bosons 
using {\tt HiggsBounds}. 
In particular, processes such as $e^+ e^- \to Z' h$, $Z'H$, 
and $H^\pm W^\mp$ are strongly suppressed 
due to the small $Z$-$Z'$ mixing angle and/or limited LEP kinematics, 
allowing the model to evade existing LEP bounds.

We note that the charged Higgs boson is also relatively light, 
with a very narrow allowed mass window: 
$110~\mathrm{GeV} < m_{H^\pm} < 170~\mathrm{GeV}$.
In this model, the charged Higgs mass is fixed by 
only the quartic coupling $\lambda_4$ 
because the $U(1)_X$ gauge symmetry forbids the $H_1^\dagger H_2$
interactions that would otherwise generate a soft mass parameter.
Once electroweak precision data and theoretical bounds such as
perturbative unitarity and vacuum stability are imposed, the scalar
potential parameters become tightly constrained, allowing only a narrow
range for $\lambda_4$.
Consequently, the charged Higgs boson is expected to be light.
Such a light charged Higgs boson serves as a promising probe 
of our model at the LHC. 
The next section is devoted to investigating the phenomenology 
of the charged Higgs boson at the LHC.

The Higgs mixing angles $\alpha$ and $\beta$, as well as the mass of the
additional neutral Higgs boson $m_h$, are not strongly constrained when $m_{Z'}
\sim 9.4~\mathrm{GeV}$.
However,
they become significantly restricted once $m_{Z'}$ exceeds approximately
$10~\mathrm{GeV}$. Notably, $\sin\alpha$ must be very small, with $|\sin\alpha|
< 0.003$ when $m_h < 62~\mathrm{GeV}$. This is because the decay channel $H \to
hh$ opens when $m_h < m_H/2 \sim 62~\mathrm{GeV}$, and the strong constraint
from the Higgs total decay width, $\Gamma_H = 3.7^{+1.9}_{-1.4}~\mathrm{MeV}$,
becomes crucial.

\section{Productions and Decays of the Charged Higgs Boson}
\label{sec:lhc}

From the allowed parameter space obtained
in the previous section,
we find that $m_{H^\pm} < 170$ GeV.
Since the charged Higgs boson is light, i.e.,
$m_{H^\pm} < m_t - m_b$,
we expect that the dominant production process at the LHC is via top quark
decay: $t \to b H^+$.
We compute the decay rate using the formula provided in
Ref.~\cite{Barger:1989rh} to obtain the branching ratio
${\rm Br}(t\to b H^\pm) =
\Gamma(t\to b H^\pm)/(\Gamma_{\rm tot}^{\rm SM}+\Gamma(t\to b H^\pm))$,
where the total width of the top quark in the SM is
$\Gamma_{\rm tot}^{\rm SM}= 1.322$~GeV~\cite{Gao:2012ja}.
We find Br$(t\to b H^\pm) < 7$\%
with the allowed parameters of the model,
which agrees with the measured value
$\Gamma_{\rm tot}^t = 1.42^{+0.19}_{-0.15}$~GeV.

\begin{figure}[t]
%\centering
%\centering
\includegraphics[width=0.46\textwidth]{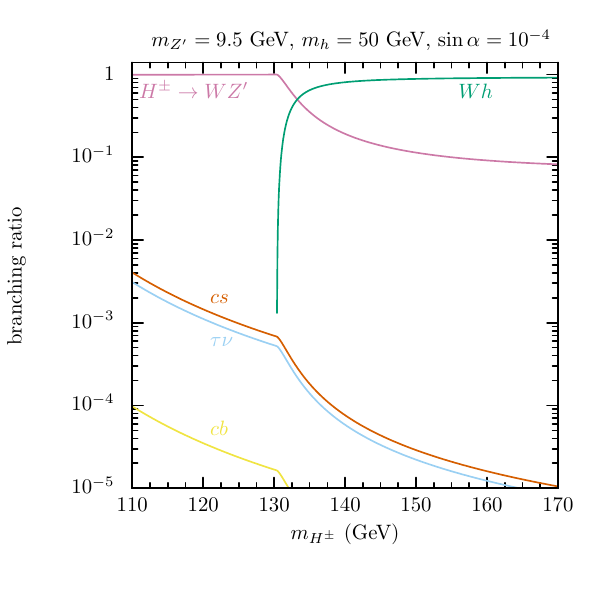}
\includegraphics[width=0.46\textwidth]{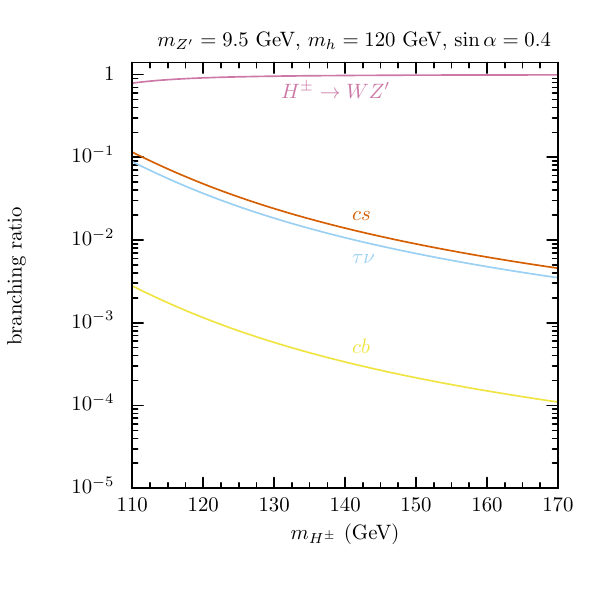}
\caption{
Branching ratios of the charged Higgs boson as functions of $m_{H^\pm}$.
Left: $m_{Z'} = 9.5~\text{GeV}$, $m_h = 50~\text{GeV}$,
and $\sin\alpha = 10^{-4}$.
Right: $m_{Z'} = 9.5~\text{GeV}$, $m_h = 120~\text{GeV}$,
and $\sin\alpha = 0.4$.
The decay modes $H^\pm \to WZ'$, $Wh$, $c\bar{s}$, $c\bar{b}$,
and $\tau\nu$ are shown.
}\label{fig:br}
\end{figure}

The produced charged Higgs boson can decay into fermion pairs, or into a $W$
boson accompanied by a neutral scalar or vector boson, if kinematically allowed.
The decay $H^\pm \to W^\pm h$ is permitted for $m_h < 90~\mathrm{GeV}$, with the
decay width given by
\be
\Gamma(H^+ \to W^+ h)
	&=& \frac{G_F \sin^2(\beta + \alpha)}{8\sqrt{2} \pi}
  m_{H^\pm}^3 \lambda^{3/2}
  \left( 1, \, m_W^2 / m_{H^{\pm}}^2, \, m_h^2 / m_{H^\pm}^2 \right) ,
\ee
where
$\lambda(a,b,c)\equiv a^2 +b^2 +c^2 -2ab -2bc -2ca$ is the usual kinematic function.
The decay width into the SM Higgs boson, $\Gamma(H^+ \to W^+ H)$, can be
obtained by replacing $\sin^2(\beta + \alpha)$ with $\cos^2(\beta + \alpha)$ and
$m_h$ with $m_H$. However, the decay $H^+ \to W^+ H$ is not allowed in this
model since $m_{H^\pm} < 170~\mathrm{GeV}$.

With the $Z/Z'$--$W^\pm$--$H^\mp$ couplings,
$H^\pm$ can also decay into $W^\pm Z/Z'$.
The decay width for $H^+ \to W^+ Z'$ is given by
\be
  \Gamma(H^+ \to W^+ Z')
  = \frac{g_{Z' W^\pm H^\mp}^2}{16 \pi m_{H^\pm}}
    \lambda^{1/2}\big(1, \, m_W^2/m_{H^\pm}^2, \, m_{Z'}^2/m_{H^\pm}^2\big)
    \left[ 2 + \frac{(m_{H^\pm}^2 - m_W^2
	     - m_{Z'}^2)^2}{4 m_W^2 m_{Z'}^2} \right] .
\ee
The decay width $\Gamma(H^+ \to W^+ Z)$ can be obtained by replacing $g_{Z'
  W^\pm H^\mp}$ with $g_{Z W^\pm H^\mp}$ and $m_{Z'}$ with $m_Z$. However, the
decay $H^+ \to W^+ Z$ is kinematically forbidden in this model.

The decays of $H^\pm$ into fermions are given by
\begin{align}
  \Gamma(H^+ \to f \bar f') =
  &~ \frac{N_c G_F}{4 \sqrt{2} \pi} \frac{m_{H^\pm}}{\tan^2\beta}
	| V_{ff'} |^2
	\lambda^{1/2} \big(1, \, m_f^2 / m_{H^\pm}^2, \,
	 m_{f'}^2 / m_{H^\pm}^2\big)
    \nonumber\\
  & \times \left[
	  \left( 1 - \frac{m_f^2}{m_{H^\pm}^2}
	           - \frac{m_{f'}^2}{m_{H^\pm}^2} \right)
	\big(m_f^2 + m_{f'}^2\big)
	+ \frac{4 m_f^2 m_{f'}^2}{m_{H^\pm}^2} \right] ,
\end{align}
where $N_c = 3$ is the color factor.
The decay widths including QCD corrections are found in
Ref.~\cite{Djouadi:2005gj}.

We depict the branching ratios of the charged Higgs boson as a function of
$m_{H^\pm}$ in Fig.~\ref{fig:br}. The bosonic decay modes, $H^\pm \to W^\pm h$
and $H^\pm \to W^\pm Z'$, dominate when they are kinematically allowed. It was
noted in Refs.~\cite{Bahl:2021str,Arhrib} that decay channels of $H^\pm$
involving the $W$ boson may play an important role in future collider
phenomenology.
The fermionic decay modes contribute less than 1\% when $\sin\alpha$ is small.
However, their branching ratios can reach up to 10\% if the $W^\pm h$ channel is
closed and $\sin\alpha$ is sizable.

\begin{figure}[t!]
\includegraphics[width=0.46\textwidth]{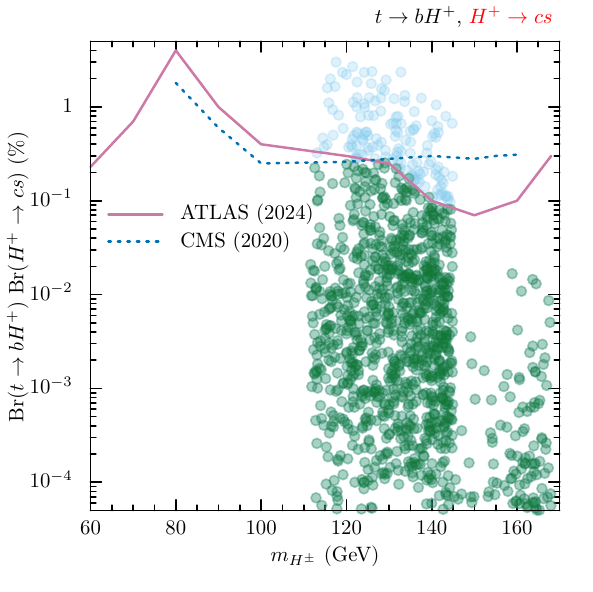}
\includegraphics[width=0.46\textwidth]{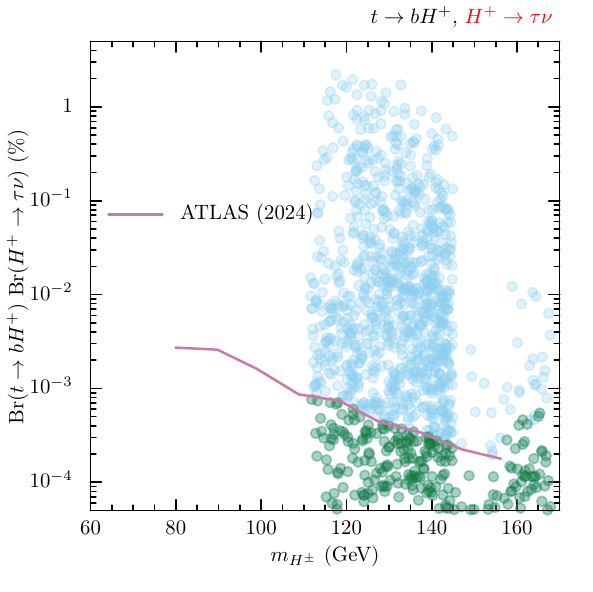}
\caption{
Model predictions for $\text{Br}(t \to H^+ b)\,\text{Br}(H^+ \to c\bar{s})$
compared with the experimental limits from the ATLAS and CMS collaborations
at the LHC~\cite{CMS:2020osd,ATLAS:2024oqu} (left),
and those for $\text{Br}(t \to H^+ b)\,\text{Br}(H^+ \to \tau\bar{\nu})$
compared with the ATLAS results~\cite{ATLAS:2024hya} (right).
}
\label{fig:light}
\end{figure}

The ATLAS and CMS collaborations have performed searches for a light charged
Higgs boson $H^\pm$ at $\sqrt{s} = 7$, 8, and 13~TeV, primarily through
fermionic decay channels. A favored search mode is $H^\pm \to cs$, which is
significant in many 2HDMs and can have a branching
ratio as high as 100\%, depending on the model. Using this decay channel,
experimental limits have been set for charged Higgs boson masses in the range
$80$--$160~\mathrm{GeV}$~\cite{CMS:2015yvc,CMS:2020osd,ATLAS:2013uxj,ATLAS:2024oqu}.
Recent searches for $H^\pm \to \tau^\pm \nu_\tau$ have also been performed by
ATLAS at $\sqrt{s} = 13~\mathrm{TeV}$~\cite{ATLAS:2024hya}. The region with
small $\sin\alpha$ is not constrained by these searches due to the suppressed
branching ratios of fermionic decays. However, if $\sin\alpha$ is sizable and
$m_h$ is large enough to close the $H^\pm \to W^\pm h$ channel, the branching
ratios $\mathrm{Br}(H^\pm \to cs)$ and $\mathrm{Br}(H^\pm \to \tau \nu)$ can
reach up to 10\%.
In such cases, recent light $H^\pm$ searches provide meaningful constraints on
the model. We present our model predictions for $\mathrm{Br}(t \to H^+
b)\mathrm{Br}(H^+ \to c\bar{s})$ and $\mathrm{Br}(t \to H^+ b)\mathrm{Br}(H^+
\to \bar{\tau} \nu)$ in Fig.~\ref{fig:light}, along with the latest ATLAS and
CMS limits from Refs.~\cite{CMS:2020osd,ATLAS:2024oqu,ATLAS:2024hya}.

As shown in Fig.~\ref{fig:br}, the branching ratios of the fermionic decays of
the charged Higgs boson are at most 10\%. Therefore, it is more promising to
investigate the bosonic decay channels $H^\pm \to W^\pm h$ and $H^\pm \to W^\pm
Z'$ for discovering the charged Higgs boson in this model.
When the final state includes the $h$ boson, its mass must satisfy $m_h <
m_{H^\pm} - m_W$, i.e., $8~\mathrm{GeV} < m_h < 90~\mathrm{GeV}$. In this mass
range, the neutral Higgs boson $h$ predominantly decays into $b\bar{b}$.
However, observing such decays into hadrons, including $h \to b\bar{b}$ and $h
\to c\bar{c}$, is challenging at the LHC due to overwhelming QCD backgrounds.
The decay $h \to \tau^+ \tau^-$ may offer a viable search channel for $h$, but
current LHC searches for new neutral Higgs bosons have focused on the heavier
mass region, $m_h > 100~\mathrm{GeV}$, which is not favored in this model.
Another opportunity to detect $h$ arises from its potentially long lifetime. The
decay rate for $h \to b\bar{b}$ is given by
\be
\Gamma(h \to b \bar{b}) = \Gamma(H \to b \bar{b}) \frac{m_h}{m_H}
	    \left( \frac{\sin \alpha}{\sin \beta} \right)^2,
\ee
which leads to the proper lifetime
\be
\tau_h \approx \tau_H \frac{m_H}{m_h} \frac{1}{\sin^2 \alpha}
       \sim \frac{10^{-21}}{\sin^2 \alpha}.
\ee
If $\sin\alpha \sim 10^{-6}$, the proper lifetime of $h$ is approximately
$10^{-9}~\mathrm{s}$, and the decay length can be $\gamma c \tau \sim
\mathcal{O}(1)~\mathrm{m}$, allowing for the possibility of observing a
displaced vertex inside the detector. For $\sin\alpha < 10^{-7}$, $h$ becomes
long-lived enough to escape the detector, potentially contributing to large
missing transverse energy, which could still serve as a detectable
signal.

\begin{figure}[t!]
\centering
\includegraphics[width=0.46\textwidth]{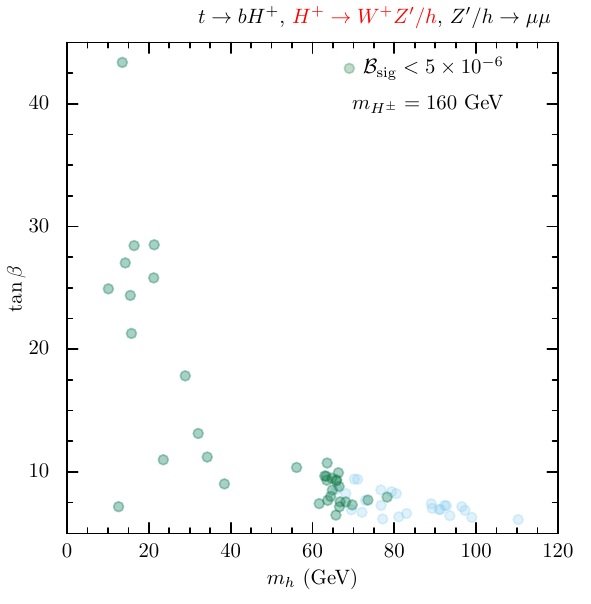}
\caption{
Parameter regions in the $(m_h, \, \tan\beta)$ plane
with and without the limits
${\cal B}_{sig} < 5\times10^{-6}$ are shown,
which are obtained by the CMS trilepton final states searches
\cite{CMS:2019idx}.
The charged Higgs mass is fixed, $m_{H^\pm} = 160~\text{GeV}$
}\label{fig:trilepton}
\end{figure}

If the DM mass is less than $m_{Z'}/2$,
$Z'$ dominantly decays into the DM fermions,
resulting in a large missing transverse energy signal.
Otherwise, $Z'$ decays into the SM fermion pairs.
However, decays into $q\bar{q}$, which produce hadronic jets, 
suffer from large backgrounds at the LHC and are difficult to identify.
A more promising strategy is to use trilepton final states, 
such as $W Z' \to e \mu \mu$ and $W Z' \to \mu \mu \mu$, 
similar to the search for a light charged Higgs boson 
decaying into a $W$ boson and a CP-odd Higgs boson $A$ in
Ref.~\cite{CMS:2019idx}. 
The CMS collaboration has presented 95\% CL
upper limits on the product of branching ratios,
${\cal B}_{\rm sig} = {\rm Br}(t \to bH^+)
{\rm Br}(H^+ \to W^+ A) {\rm Br}(A \to \mu^- \mu^+))$
based on a combined likelihood analysis of event yields in the $e\mu\mu$ and
$\mu\mu\mu$ channels. These upper limits are not strongly sensitive to $m_A$.
To apply these constraints to our model, we replace $A$ with $Z'$ or $h$ and
evaluate $\mathcal{B}_{\mathrm{sig}}$ for $m_{H^\pm} = 160~\mathrm{GeV}$. 
The parameter space excluded by the CMS trilepton constraints 
corresponds to regions with large $m_h$ and small $\tan\beta$, 
as shown in Fig.~\ref{fig:trilepton}.
Small $\tan\beta$ enhances the coupling $g_{Z' H^\pm W^\mp}$, 
leading to a large branching ratio $\mathrm{Br}(H^+ \to W^+ Z')$.
On the other hand, the region with $m_h > 80~\mathrm{GeV}$ is kinematically
disallowed since $m_{H^\pm}$ is fixed to $160~\mathrm{GeV}$.

Since the dark $Z$ boson is light, with a mass around $10~\mathrm{GeV}$, it is
expected to be highly boosted at the LHC. The light $Z'$ decays into lighter
SM particles, such as electron pairs, muon pairs, and charged pion
pairs. These decay products, being produced with large boosts, form clusters of
highly collimated leptons that resemble jet-like structures, commonly referred
to as lepton jets.
Such clusters of energetic and collimated leptons are distinctive signatures at
the LHC. Collider signatures of lepton jets have been studied in
Refs.~\cite{Falkowski:2010gv,Arkani-Hamed:2008kxc}. 
Therefore, charged Higgs boson production in this model 
offers an opportunity to investigate lepton jets,
although we do not explore this aspect further in the present work.

\begin{figure}[t]
\centering
\includegraphics[width=10cm]{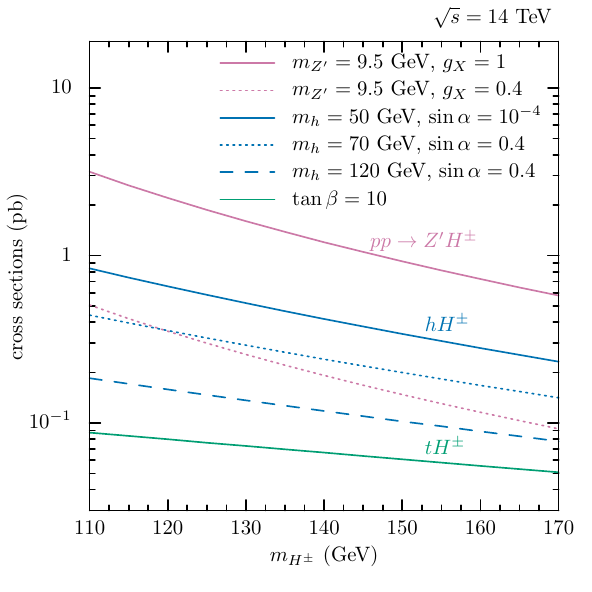}
\caption{
Production cross sections of the charged Higgs boson in association with
the dark $Z$, the neutral Higgs $h$, and the top quark.
The results are shown as functions of $m_{H^\pm}$
for representative parameter sets:
$m_{Z'} = 9.5~\text{GeV}$ with $g_X = 1$ and $0.4$,
and $m_h = 50$, 70, 120~GeV with $\sin\alpha = 10^{-4}$ and $0.4$.
The value of $\tan\beta$ is fixed at 10.
}\label{fig:production}
\end{figure}

In addition to top-quark decays, 
there exist direct production processes of the charged Higgs boson at the LHC. 
One of the most important channels is its
associated production with top quarks:
\be
pp \to g b \to t H^\pm,
\ee
driven by the large Yukawa coupling of the top quark.
In this model, substantial production of $H^\pm$ is also expected 
in association with either the dark $Z$ boson or 
the neutral Higgs boson $h$ via $W^\pm$ exchange:
\be
pp \to W^\pm \to H^\pm Z',~~H^\pm h.
\ee
We present the direct production cross sections for $H^\pm Z'$, $H^\pm h$, and
$H^\pm t$ in Fig.~\ref{fig:production}. 
These processes can reach cross sections
of $\sigma(pp \to H^\pm Z'/h) \sim \mathcal{O}(1)~\mathrm{pb}$ within the
allowed parameter space.
Assuming an integrated luminosity of $250~\mathrm{fb}^{-1}$ during LHC Run 3, 
we expect that more than $10^5$ charged Higgs bosons 
could be directly produced in the most optimistic scenario.

One of the characteristic signals for direct $H^\pm$ production is the
observation of two lepton jets emitted in opposite directions: one originating
from the associated production and the other from the decay of $H^\pm$. If
$H^\pm$ is produced in association with $Z'$ or $h$, a promising final state is
a five-muon signal, $\mu\mu\mu\mu\mu$, arising from $Z' \to \mu^+\mu^-$ or $h
\to \mu^+\mu^-$, along with the muonic decay of the $W$ boson:
\be
p p \to H^\pm Z' / h \to (W^\pm Z' /h) Z' / h \to
(\mu \nu)(\mu \mu)(\mu \mu)~.
\ee
For example, assuming $m_{Z'} = 9.5~\mathrm{GeV}$, 
$m_{H^\pm} = 120~\mathrm{GeV}$, 
and an integrated luminosity of $250~\mathrm{fb}^{-1}$ during LHC Run 3, 
we estimate:
\begin{itemize}
  \item $\sigma(pp \to H^\pm Z') \approx 1~\mathrm{pb}$,
  \item $\mathrm{Br}(Z' \to \mu^+ \mu^-) \sim 10\%$,
  \item $\mathrm{Br}(W \to \mu \nu) \sim 10\%$,
\end{itemize}
which result in approximately $500$ five-muon events.
Such multi-muon final states provide a clean and distinctive signature at the
LHC, offering a promising avenue for probing the charged Higgs boson in this
model.

\section{Dark Matter Phenomenology}
\label{sec:dm}

 \begin{figure}[t]
 \centering
 \includegraphics[width=15cm]{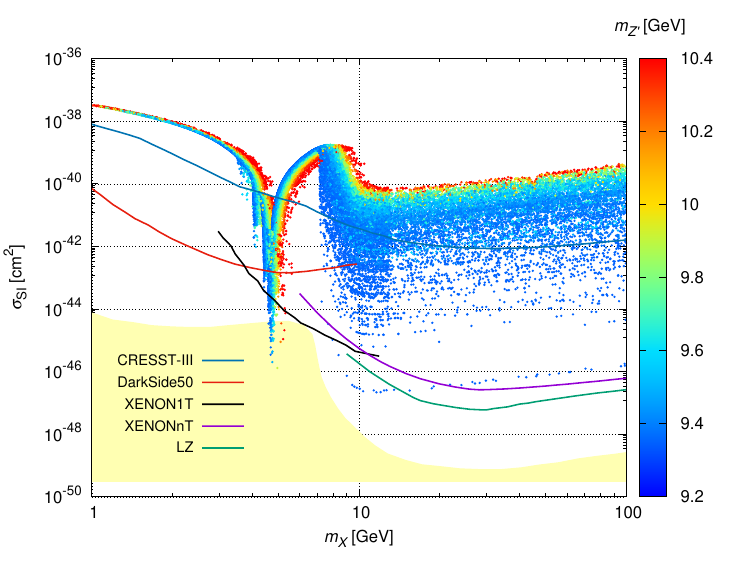}
 \caption{The DM–nucleon cross sections corresponding to the parameter sets that yield the relic abundance within 
 the 3$\sigma$ range are shown. Constraints from the Bullet Cluster on DM self-interactions are also taken into account.}
 \label{fig:DMDD}
 \end{figure}

In this section, we briefly discuss the DM phenomenology
within the benchmark scenarios relevant for charged Higgs boson searches.
The hidden sector Lagrangian describing DM interactions takes a QED-like form:
\be
{\cal L}_{\text{hidden}} = -\frac{1}{4} F_X^{\mu \nu} F_{X \mu \nu}
                    + \bar{\psi}_X i \gamma^\mu D_\mu \psi_X
                          - m_X \bar{\psi}_X \psi_X,
\ee
where
\be
D^\mu = \partial^\mu + i g_X A^\mu_X X.
\ee
The U(1)$_X$ symmetry is broken by EWSB,
and the DM fermion $\psi_X$ communicates with the SM
through the $Z$ and dark $Z$ bosons.
The fermion mass $m_X$ and the gauge charge $X$ are
not constrained by any of the observables discussed in Sec.~\ref{sec:model}.

From the Higgs phenomenology studied in the previous sections,
including the latest ATLAS and CMS results 
on charged Higgs boson production and decay,
we obtain stringent bounds on the dark $Z$ and charged Higgs boson masses: 
$9.3$~GeV$\lesssim m_{Z'} \lesssim 11.3$~GeV, 
and $ 110~{\rm GeV}\lesssim m_{H^\pm} \lesssim 170~{\rm GeV}$.
Since the DM properties are mostly independent of the charged Higgs mass, 
we focus on the DM phenomenology driven by the dark $Z$ mass.

The dark gauge coupling $g_X$ is also tightly constrained
by electroweak observables and Higgs data,
as discussed in Secs.~\ref{sec:model}--\ref{sec:lhc}.
Nevertheless, the DM interaction allows a somewhat wider parameter space
when varying the DM charge $X$ under the U(1)$_X$ symmetry. 
In this case, the coupling between the DM and the dark $Z$
is given by $g_X X$.

Adopting the standard thermal freeze-out scenario, 
we investigate the parameter sets that reproduce the observed
relic density using ${\tt micrOmegas}$~\cite{Alguero:2023zol}. 
Our analysis shows that either a sharp resonance condition
$2 m_X \sim m_{Z'}$ or a moderately heavier DM 
with $m_X$ near $m_{Z'}$ or larger is required to evade the various
current direct detection bounds.
As a consequence, the allowed DM mass is restricted to narrow ranges, 
around $4.5$--$5.5$~GeV or $8$--$14$~GeV 
as shown in Fig.~\ref{fig:DMDD}.
This indicates that a certain degree of tuning is needed
to simultaneously satisfy the relic density and experimental constraints.
A more natural explanation may require extensions of the dark sector
and/or alternative mechanisms for generating the DM relic abundance
beyond the conventional freeze-out scenario.
We leave a detailed exploration of these possibilities for future work.

We also need to discuss constraints from DM annihilation in a later time 
such as indirect searches~\cite{Fermi-LAT:2015att} 
or CMB limits~\cite{Slatyer:2015jla}.
The allowed region in Fig.~\ref{fig:DMDD} consists of two parts: 
(1) resonance region $m_X\lesssim m_{Z'}/2$, and 
(2) $Z'$ threshold region, $m_X\sim m_{Z'}$.
In region (1), the dominant annihilation is 
$\psi_X \bar\psi_X \to Z^{\prime (\ast)} \to f \bar f$, 
where $f$ denotes SM particles.
In this case, the correct relic density is determined 
by the resonantly enhanced annihilation with the help of
sizable DM velocity, $v_{\psi_X} \sim 1/3$.
At a lower temperature, the same annihilation is off the resonance peak, 
so becomes ineffective.
On the other hand, in region (2), 
the dominant annihilation channel is $\psi_X \bar\psi_X \to Z' Z'$. 
Especially for $m_X \lesssim m_{Z'}$, 
this process is open only when the DM has a substantial kinetic energy, 
{\it i.e.} at a high temperature~\cite{Griest:1990kh}.
At a lower temperature, however, this channel is closed 
for most of the DM particles and thus the annihilation process 
becomes irrelevant.
For $m_X>m_{Z'}$, this argument does not hold and 
the model is indeed constrained.
We have numerically checked these observations using {\tt micrOmegas}.
The averaged annihilation cross section well after the freeze-out, 
$\langle \sigma v\rangle$ in region (1) and part of region (2)
is smaller than $10^{-27}~\text{cm}^2$ 
while some allowed parameter points in Fig.~\ref{fig:DMDD} give 
larger values and are further constrained.
However, the generic feature of DM phenomenology does not alter.

\section{Concluding Remarks}
\label{sec:conc}

Searches for charged Higgs bosons at the LHC have been presented in the context
of a dark $Z$-mediated fermionic DM model. We propose several search
channels for $H^\pm$ produced via top-quark decays, depending on the model
parameters. Trilepton final states such as $H^\pm \to W^\pm Z' \to e\mu\mu$ or
$\mu\mu\mu$ may be promising if the DM fermion mass exceeds half the
$Z'$ boson mass.
When kinematically allowed, the decay $H^\pm \to W^\pm h$ becomes dominant, and
the subsequent decay $h \to \tau^+ \tau^-$ is accessible for $m_h <
90~\mathrm{GeV}$. If the neutral Higgs mixing angle $\alpha$ is sufficiently
small, displaced vertices from $h \to f\bar{f}$ decays could serve as
distinctive signals.
The conventional search channels $H^\pm \to cs$ and $H^\pm \to \tau \nu$ still
offer potential for discovering the charged Higgs boson. Additionally, direct
production of $H^\pm$ at the LHC may yield conspicuous five-muon final states.
Even during LHC Run 3, we expect up to 500 events with 
$(\mu\mu\mu\mu\mu)$ final
states.

We verify that the model parameter space allowed by recent ATLAS and CMS 
charged Higgs searches is consistent with the observed relic abundance 
from cosmic microwave background measurements 
via thermal freeze-out, and also satisfies direct detection bounds. 
As a result, our model not only features a viable fermionic DM candidate, 
but also offers rich and testable
phenomenology for the charged Higgs boson at the LHC.

\acknowledgments

This work is supported
by Basic Science Research Program
through the National Research Foundation of Korea
funded by the Ministry of Science and ICT
under the Grants
RS-2022-NR070836 (KJB),
RS-2024-00336507 (JG),
RS-2021-NR059413 (KYL, CY),
RS-2023-00237615 (CY),
and RS-2023-00209974 (CBP).
DWJ is supported by IBS under the project code IBS-R018-D3.
JG also acknowledges the Ewha Womans University
Research Grant of 2024 (1-2024-0651-001-1).
JG is grateful to the Asia Pacific Center for Theoretical Physics
for hospitality while this work was in progress.

\section*{Appendix}
\label{sec:couplings}

We here present the couplings of the charged Higgs boson that are relevant for
its production and decay processes.
The couplings of the charged Higgs boson to SM fermions are given by
\be
g_{H^- \bar{f}f'}= \frac{\sqrt{2}}{v \tan \beta}
               \left( m_f P_L -m_{f'} P_R \right),
\ee
which are proportional to $1/\tan \beta$.
In contrast, in the Type-II 2HDM,
the couplings are proportional to either
$\tan \beta$ and $1/\tan \beta$ depending on the fermion's weak isospin.
Due to these coupling structures, the constraint from
the $b \to s \gamma$ constraint requires
$\tan \beta$ be rather large, specifically $\tan \beta > 3$,
unless $m_{H^\pm}$ is very large.

The $H^\pm$ couplings to neutral gauge bosons are given by
\be
g_{Z H^\pm H^\mp} &=& \frac{1}{c_W} \left[ (1-2s_W^2) c_X
              + \frac{g_X}{g} c_W s_X \sin^2 \beta \right] , ~~~
\nonumber  \\
g_{Z' H^\pm H^\mp} &=& \frac{1}{c_W} \left[ -(1-2s_W^2) s_X
              + \frac{g_X}{g} c_W c_X \sin^2 \beta \right],
\ee
where the U(1)$_X$ gauge coupling $g_X$ is expressed by
\be
\frac{g_X}{g} = -\frac{s_X c_X c_W}{ \cos^2 \beta}
                 \frac{m_Z^2-m_{Z'}^2}{m_W^2}.
\ee
We see that
$g_{Z H^\pm H^\mp} = (1-2s_W^2)/c_W + {\cal O}(s_X^2)$
and $g_{Z' H^\pm H^\mp} = {\cal O}(s_X)$,
which are close to the corresponding couplings in the typical 2HDM,
$g_{Z H^\pm H^\mp} = (1-2s_W^2)/c_W$,
in the limit $s_X \to 0$.

The $H/h$--$H^\pm$--$W^\pm$ boson couplings are given by
\be
g_{H H^\pm W^\mp} &=& \pm \frac{e}{2 s_W} \cos(\alpha+\beta),~~~
\nonumber \\
g_{h H^\pm W^\mp} &=& \mp \frac{e}{2 s_W} \sin(\alpha+\beta).
\ee
From Table~\ref{tab1}, we have $\alpha \ll 1$ and $\sin \beta \approx 1$,
which implies $\sin(\alpha+\beta) \approx 1$
and $\cos(\alpha+\beta) \approx 0$.
Hence, $g_{h H^\pm W^\mp}$ is substantial
without $s_X$ suppression, while
$g_{H H^\pm W^\mp}$ is very small.

The $Z/Z'$--$H^\pm$--$W^\mp$ couplings given by
\be
g_{Z H^\pm W^\mp} &=& - g_X m_W \cos \beta \sin \beta s_X, ~~~
\nonumber \\
g_{Z' H^\pm W^\mp} &=& - g_X m_W \cos \beta \sin \beta c_X.
\ee
In typical 2HDMs without U(1)$_X$,
there are no $Z/Z'$--$H^\pm$--$W^\mp$ couplings.
Instead, an $A$--$H^\pm$--$W^\mp$ coupling exists.
Since the CP-odd scalar $A$ becomes the longitudinal mode
of the extra neutral massive gauge boson in our model,
the $A$--$H^\pm$--$W^\mp$ coupling transitions into
the $Z'$--$H^\pm$--$W^\mp$ coupling, and also contributes to
the $Z$--$H^\pm$--$W^\mp$ coupling via mixing.
Thus, the existence of the $Z/Z'$--$H^\pm$--$W^\mp$ couplings
can serve as a distinctive signature of new physics beyond the
2HDM~\cite{Ko:2012hd}.
It should be noted that
the $Z-H^\pm-W^\mp$ coupling is highly suppressed
by both $s_X$ and $\cos \beta$.

Finally, the Higgs triple couplings with the charged Higgs are
listed below:
\be
g_{HH^+ H^-} &=& \frac{v^2}{m_Z^2} \Big[
             2 \lambda_1 \sin \alpha \cos \beta \sin^2 \beta
                  + 2 \lambda_2 \cos \alpha \sin \beta \cos^2 \beta
      + \lambda_3 (\cos \alpha \sin^3 \beta+\sin \alpha \cos^3 \beta)
\nonumber \\
&&~~~~~~~  - \lambda_4 (\cos \alpha \cos^2 \beta \sin \beta
                             +\sin \alpha \cos \beta \sin^2 \beta)
      \Big] ,
\nonumber \\
g_{hH^+ H^-} &=& \frac{v^2}{m_Z^2} \Big[
      2 \lambda_1 \cos \alpha \cos \beta \sin^2 \beta
                  - 2 \lambda_2 \sin \alpha \sin \beta \cos^2 \beta
      + \lambda_3 (\cos \alpha \cos^3 \beta-\sin \alpha \sin^3 \beta)
\nonumber \\
&&~~~~~~~  + \lambda_4 (\sin \alpha \cos^2 \beta \sin \beta
                             -\cos \alpha \cos \beta \sin^2 \beta)
      \Big],
\ee
which are not relevant for our analysis.

%%%%%%%%%%%%%%%%%% References
%%%%%%%%%%%%%%%%%%%%%%%%%%%%%%%%%%%%%%%%%%%%%%%%%%%%%%%
\def\PRDD #1 #2 #3 {Phys. Rev. D \textbf{#1},\ #2 (#3)}
\def\PRD #1 #2 #3 #4 {Phys. Rev. D \textbf{#1},\ No. #2, #3 (#4)}
\def\PRLL #1 #2 #3 {Phys. Rev. Lett. {\bf#1},\ #2 (#3)}
\def\PRL #1 #2 #3 #4 {Phys. Rev. Lett. {\bf#1},\ No. #2, #3 (#4)}
\def\PRA #1 #2 #3 {Phys. Rev. A {\bf#1},\ #2 (#3)}
\def\PLB #1 #2 #3 {Phys. Lett. B {\bf#1},\ #2 (#3)}
\def\NPB #1 #2 #3 {Nucl. Phys. B {\bf #1},\ #2 (#3)}
\def\ZPC #1 #2 #3 {Z. Phys. C {\bf#1},\ #2 (#3)}
\def\EPJ #1 #2 #3 {Euro. Phys. J. C {\bf#1},\ #2 (#3)}
\def\JPG #1 #2 #3 {J. Phys. G: Nucl. Part. Phys. {\bf#1},\ #2 (#3)}
\def\JHEP #1 #2 #3 {JHEP {\bf#1},\ #2 (#3)}
\def\JCAP #1 #2 #3 {JCAP {\bf#1},\ #2 (#3)}
\def\IJMP #1 #2 #3 {Int. J. Mod. Phys. A {\bf#1},\ #2 (#3)}
\def\MPL #1 #2 #3 {Mod. Phys. Lett. A {\bf#1},\ #2 (#3)}
\def\PTP #1 #2 #3 {Prog. Theor. Phys. {\bf#1},\ #2 (#3)}
\def\PR #1 #2 #3 {Phys. Rep. {\bf#1},\ #2 (#3)}
\def\RMP #1 #2 #3 {Rev. Mod. Phys. {\bf#1},\ #2 (#3)}
\def\PRold #1 #2 #3 {Phys. Rev. {\bf#1},\ #2 (#3)}
\def\IBID #1 #2 #3 {{\it ibid.} {\bf#1},\ #2 (#3)}
\def\etal {{\it et al.} }

\end{document}